\documentclass[fleqn,twoside,twocolumn,nofootinbib,showkeys]{revtex4} 
\usepackage[nocpr,nopacs]{ujp} 

\usepackage{color, graphicx}
\begin{document}
\title[Method of X-ray Diffraction Data Processing]
{METHOD OF X-RAY DIFFRACTION\\ DATA PROCESSING FOR MULTIPHASE\\
MATERIALS WITH LOW PHASE CONTENTS}
\author{A.D.~Skorbun}
\affiliation{Institute for Safety Problems of Nuclear Power Plants,
Nat. Acad. of Sci. of Ukraine}
\address{36a, Kirov Str., Chornobyl 07270, Kyiv
region, Ukraine}
\email{i.zhyganiuk@ispnpp.kiev.ua}
\author{S.V.~Gabielkov}
\affiliation{Institute for Safety Problems of Nuclear Power Plants,
Nat. Acad. of Sci. of Ukraine}
\address{36a, Kirov Str., Chornobyl 07270, Kyiv
region, Ukraine}
\email{i.zhyganiuk@ispnpp.kiev.ua}
\author{I.V.~Zhyganiuk}
\affiliation{Institute for Safety Problems of Nuclear Power Plants,
Nat. Acad. of Sci. of Ukraine}
\address{36a, Kirov Str., Chornobyl 07270, Kyiv
region, Ukraine}
\email{i.zhyganiuk@ispnpp.kiev.ua}
\author{V.G.~Kudlai}
\affiliation{Institute for Safety Problems of Nuclear Power Plants,
Nat. Acad. of Sci. of Ukraine}
\address{36a, Kirov Str., Chornobyl 07270, Kyiv
region, Ukraine}
\author{P.E.~Parkhomchuk}
\affiliation{Institute for Safety Problems of Nuclear Power Plants,
Nat. Acad. of Sci. of Ukraine}
\address{36a, Kirov Str., Chornobyl 07270, Kyiv
region, Ukraine}
\author{S.A.~Chikolovets}
\affiliation{Institute for Safety Problems of Nuclear Power Plants,
Nat. Acad. of Sci. of Ukraine}
\address{36a, Kirov Str., Chornobyl 07270, Kyiv
region, Ukraine}
\udk{???} \razd{\secx}

\autorcol{A.D.\hspace*{0.7mm}Skorbun, S.V.\hspace*{0.7mm}Gabielkov,
I.V.\hspace*{0.7mm}Zhyganiuk et al.}

\setcounter{page}{870}%

\begin{abstract}
Amorphous, glass, and glass-ceramic materials practically always
include a significant number (more than eight) of crystalline
phases, with the contents of the latter ranging from a few wt.\% to
several hundredths or tenths of wt.\%.\,\,The study of such
materials using the method of X-ray phase analysis faces
difficulties, when determining the phase structure.\,\,In this work,
we will develop a method for the analysis of the diffraction
patterns of such materials, when diffraction patterns include X-ray
lines, whose intensities are at the noise level.\,\,The
identification of lines is based on the search for correlations
between the experimental and test lines and the verification of the
coincidence making use of statistical methods (computer
sta\-tis\-tics).\,\,The method is tested on the specimens of
$\alpha$-quartz, which are often used as standard ones, and applied
to analyze lava-like fuel-containing materials from the destroyed
Chornobyl NPP Unit~4.\,\,It is shown that the developed technique
allows X-ray lines to be identified, if the contents of separate
phases is not less than 0.1~wt.\%.\,\,The method also significantly
enhances a capability to determine the phase contents quantitatively
on the basis of lines with low intensities.
\end{abstract}
\maketitle

\section{Introduction}

As a result of the accident at the Chornobyl NPP, the main part of
the nuclear fuel has spread over the premises of Unit~4 in the form
of a melt.\,\,These lava-like fuel-containing materials (LFCMs)
include the main part of radionuclides of the spent fuel and,
therefore, determine the nuclear, radiation, and environmental
safety of the complex \textquotedblleft New Safe Confinement --
Shelter Object\textquotedblright\ \citep{Arutyunyan}.\,\,In order to
predict a change in the state of fuel-containing materials with
time, it is necessary to know their structural parameters, including
their type, size, and the contents of crystalline
phases~\citep{1Gabielkov}.

When applying the X-ray phase analysis to them, it was found that,
firstly, the LLFCM specimens, in addition to glass on the basis of
the silicon, uranium, aluminum, and zirconium oxides, contain also
many (more than eight) crystalline phases; secondly, since the
contents of those phases in the material are very low (from a few
wt.\% to several hundredths or tenths of wt.\%), the intensities of
a many hypothetically possible reflections (X-ray lines) from them
are at the noise level; and thirdly, the number of such
low-intensity reflections becomes extremely large: up to one or two
hundred reflections \citep{GabielkovZhyganiuk}.\,\,The application of
a specialized software \citep{3Match} in order to interpret such
\textquotedblleft noisy\textquotedblright\ diffraction patterns
brings about a high degree of uncertainty.\,\,For instance, the
analysis can result in the resolution of several dozens of X-ray
lines that often correspond to compounds, whose presence in LLFCM
specimens cannot be imagined even \mbox{hypothetically.}

Hence, in order to interpret such X-ray diffraction patterns, where
the intensities of true lines are at the noise level and which
include X-ray lines from unknown phases, it is necessary to apply
methods that allow weak lines to be detected against the noise
background.\,\,In this work, a corresponding method for detecting
low-intensity X-ray lines is developed on the basis of correlation
analysis.\,\,Its efficiency and capabilities are analyzed using
X-ray diffraction data obtained from reference specimens of
$\alpha$-quartz.\,\,The application of the developed method in order
to analyze the composition and the content of crystalline phases in
LLFCMs will be demonstrated.

When discussing the results obtained, the main attention is focused
on comparing the line intensities from our experimental X-ray
diffraction data and the corresponding line intensities from the
Crystallography Open Database (COD) \citep{COD}.\,\,The\-refore, the
relative units for intensities are used, i.e.\,\,all X-ray
diffraction data are normalized to the maximum intensity value for a
given specimen or phase and multiplied by 1000.\,\,The renormalized
quantities are marked by the tilde.\,\,For instance, $\widetilde{L}%
=1000\,L/L_{\max}$.

\section{Experimental Materials}

The sequence of works is as follows.\,\,At the first stage, the
developed method was applied to analyze the diffraction patterns
from a reference specimen of $\alpha$-quartz, with its X-ray
characteristics being well known \citep{COD}.\,\,Then, after
determining the capabilities of the method, it was applied to
analyze X-ray diffraction data obtained from LLFCM specimens (the
so-called brown ceramics).

It should be emphasized that LLFCM diffraction patterns look like
noise tracks with single low-intensity lines.\,\,Ma\-king no use of
special analysis, those diffraction patterns cannot be used for the
identification of low-content phases even with the help of a
specialized software \citep{3Match}.

\section{Experimental Technique}

The phase composition of the examined materials was determined using
the X-ray diffraction method on a DRON-4 diffractometer (the
$\theta-\theta$ scheme, Cu~K$_{\alpha}$ radiation).\,\,In view of
the high radioactivity of researched materials, a system of lead
screens was installed in order to protect the personnel from
$\gamma$-radiation emitted by the specimens.\,\,A protective lead
screen was also installed to protect the photoelectron multiplier in
the diffractometer and the monochromator (a graphite crystal) in
order to reduce the influence of $\gamma$-radiation emitted by the
specimens on the useful signal.\,\,To evaluate the content of
uranium oxide, a specimen of non-irradiated nuclear fuel, which was
represented by uranium oxide UO$_{2}$, was used as a reference one.

\section{Correlation Method\\ for Signal Resolution Against\\ the Noise
Background}

In a standard automated experiment, the diffraction patterns are
obtained as discrete sets of numbers (line intensities) measured at
every angle.\,\,The positions of X-ray lines in the diffraction
pattern (the angle value 2$\theta $) are compared with the angles
calculated using the data taken from the COD database
\citep{COD}.\,\,This procedure is carried out for each analyzed
compound.

The developed correlation method consists in calculating the
correlation between a certain test line with known parameters and a
corresponding section in the diffraction pattern.\,\,Then the degree
of correlation is additionally analyzed using statistical
methods.\,\,This method was used for the first time in work
\citep{Panasyuk}, while resolving low-intensity lines from noise in
gamma spectra.

A test line, as well as a section of diffraction pattern selected
for the comparison, is a set of consecutive numbers.\,\,In terms of
statistical analysis, those sets of numbers are samples, between
which the degree of correlation is calculated.\,\,But below, to
clarify the situation, the term the correlation between the
diffraction pattern and the test line will be used.

\section{Calculation Technique}

When using the automated method for finding a useful signal in the
diffraction pattern, certain search criteria have to be set.\,\,The
following general assumptions form their basis:

\raisebox{0.5mm}[0cm][0cm]{{\footnotesize$\bullet$}}\,\,each X-ray
line is bell-shaped;

\raisebox{0.5mm}[0cm][0cm]{{\footnotesize$\bullet$}}\,\,in this
work, pseudorandomly or normally distributed samples were taken for
the theoretical analysis of the background, although any assumption
about the corresponding distribution form was not used in real
calculations.

When the noise magnitude becomes comparable with the intensity of a
useful signal, the presence of a line in the pattern cannot be
detected visually.\,\,In such cases, there arises an issue
concerning the reliability degree of the decision about the presence
of a line in the diffraction pattern.\,\,This parameter can be
estimated making use of a special statistical \mbox{treatment.}

The choice of the angle value corresponding to the line maximum in
the proposed method was made with the help of computer statistics
\citep{Moore}.\,\,The numerical data were analyzed using the
statistical permutation test, i.e.\,\,a correlation between the test
and examined lines was found.

\section{Method of Correlation Analysis}

\begin{figure}
\vskip1mm
\includegraphics[width=\column]{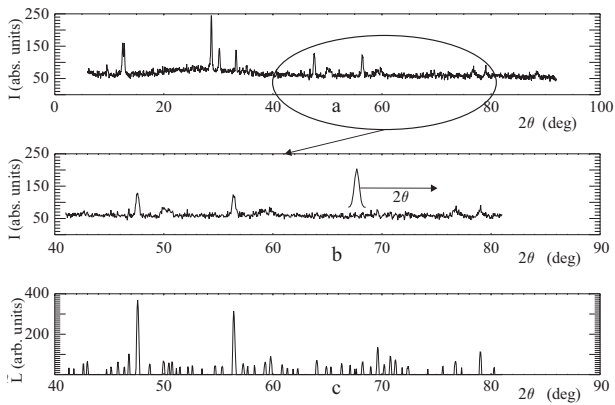}
\vskip-3mm\caption{ General appearance of the experimental
diffraction pattern ($a$), a fragment of this diffraction pattern
with the test line ($b$), and the corresponding correlation
diffraction pattern~($c$) }\label{f1}\vspace*{-2mm}
\end{figure}

In Fig.~\ref{f1}, a general schematic diagram of the method is illustrated.
Any diffraction pattern is considered as a sample of $n$ elements,
where $n$ is the number of scanning angles.\,\,In the diffraction
pattern, an interval (window) of $m$ angles is selected, whose width
corresponds to the line width, and whose center is in the middle of
the selected $m$-interval in the $k$-th angle of a diffraction
pattern.\,\,The number of impulses (the intensity) at each angle
from this window, $y_{i}$ ($i=1,2,...,k_{0},...,m$), forms a sample
for the further analysis.\,\,Then a model sample $z_{i}$ consisting
of $m$ elements (angles) is generated in a way to form a bell-shaped
line: the $z_{i}$-values are maximum in the middle of the sample and
decrease symmetrically toward the edges.\,\,In this work, a Gaussian
function, whose shape is close to the shape of a line in the
diffraction pattern, was taken for the sake of generation simplicity
(see Fig.~\ref{f1},~$b$).\,\,For those two samples, the correlation is
calculated using the permutation test.\,\,It is obvious that the
correlation will be maximum, if the centers of those lines
coincide.\,\,As a result of calculations, a histogram of the
possible values for the correlation coefficient is obtained.\,\,The
value at the histogram maximum is taken as the value of the
correlation coefficient $c_{k}$ and assigned to the $k_{0}$-th
element of a new sample of coefficients $c$.\,\,Then the window is
shifted to the right by one step, the procedure is repeated, and the
element $c_{k+1}$ is calculated.\,\,As a result of the complete scan
of the diffraction pattern, a new sample is formed from the elements
$c_{k}$ ($k=1,2,...,k_{\max}$), which will be called the correlation
diffraction pattern (see Fig.~\ref{f1},~$c$).

The shape of experimental lines is not perfect, especially at low
intensities and in the presence of the noise component.\,\,This
means that the shape is not reproduced exactly at repeated
measurements.\,\,The uncertainty degree for the measured intensity
(the area)~-- under bad conditions, the angle of maximum intensity
may also become uncertain~-- can be estimated using a special
statistical analysis, namely, by applying the permutation test
\citep{Moore}.

The essence of the latter is as follows.\,\,Let us have a sequence
of $n$ angles: $x_{1},x_{2},...,x_{0},...,x_{i},...,x_{n}$.\,\,Then
let us define a line as the number of impulses at the $i$-th
angle,\vspace*{-2mm}
\begin{equation}
y_{i}=A_{0}\exp\left[  -\frac{{(x_{i}-x_{0})^{2}}}{{2\sigma^{2}}}\right]
+\varepsilon_{i},\label{1}%
\end{equation}
where $\sigma^{2}$ is the dispersion, $x_{0}$ the position of the
curve maximum, and $\varepsilon_{i}$ the random component.\,\,We
define the test
line as the Gaussian function%
\begin{equation}
z_{k}=A_{1}\exp\left[
-\frac{{(x_{k}-x_{c})^{2}}}{{2\sigma^{2}}}\right]\!\!
,\label{2}%
\end{equation}
where $k=1,2,...,m$; $c=m/2$ is the sample middlepoint, and
$m\approx \sigma\leq n$.\,\,Now, let us choose $m$ neighbor elements
from the sequence
$y_{i}$~-- i.e. $y_{i},...,y_{i+m}$~-- and form the sum%
\begin{equation}
S_{j0}=\sum\limits_{k=j}^{m}{y_{k}z_{k}}.\label{3}%
\end{equation}
It is easy to verify that this quantity is maximum, when
$x_{c}=x_{0}$.\,\,Pro\-vi\-ded that the set $z_{k}$ is fixed, let us
randomly rearrange the elements in the sample $y_{k}$ and calculate
a new sum $S_{j}$.\,\,If this procedure is repeated many times, then
the percentage $P$ of $S_{j}$-va\-lues exceeding $S_{j0}$ will
correspond to the probability that the initial sum $S_{j0}$ arose
owing to a random relationship between $y_{k}$ and $z_{k}$ (this is
the so-called $P$-cri\-te\-rion \citep{Moore}).\,\,For convenience,
the reference point
can be shifted to zero, thus introducing the quantity $\Delta S_{j}%
=S_{j0}-S_{j}$.\,\,Then the negative values of $\Delta S$ testify
that a correlation may appear at random $y_{k}$-realizations.\,\,The
deviation $L_{j}$ of the maximum in the histogram of the quantities
$\Delta S_{j}$ from zero is selected as the correlation magnitude.

One of the key issues of the method developed to find low-intensity
lines is the question of how the proposed algorithm treats the
background, i.e.\,\,which background is in the correlation
diffraction pattern.\,\,To answer it, a diffraction pattern was
registered in the absence of the specimen and processed making use
of the described correlation method.\,\,In Fig.~\ref{f2}, the result of
such a processing of the background signal with no real lines in the
diffraction pattern is demonstrated.\,\,Let the notation $\delta$
stand for the background level in the X-ray diffraction
data.\,\,Since there are no real lines in the examined signal, a
conclusion is drawn that, for the selected processing mode, the
background corresponds to the parameter value $\delta
\leq4$.\,\,This means that the discrimination level at the
processing of real diffraction patterns has to be not be less than
four~intensity units in the correlation diffraction pattern.\,\,In
other words, all lines in the correlation diffraction pattern with
the intensity less than or equal to 4 should be put equal
to~zero.\vspace*{-2mm}

\section{Verification of the Method\\ on X-ray Diffraction Data Obtained\\ for
Reference Specimens}

\begin{figure}
\vskip1mm
\includegraphics[width=\column]{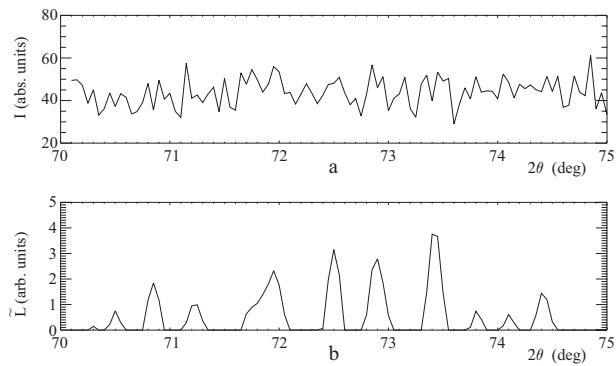}
\vskip-3mm\caption{ Diffraction pattern obtained without a specimen,
i.e. the background ($a$); its correlation diffraction pattern ($b$)
}\label{f2}
\end{figure}

\begin{figure}
\vskip3mm
\includegraphics[width=\column]{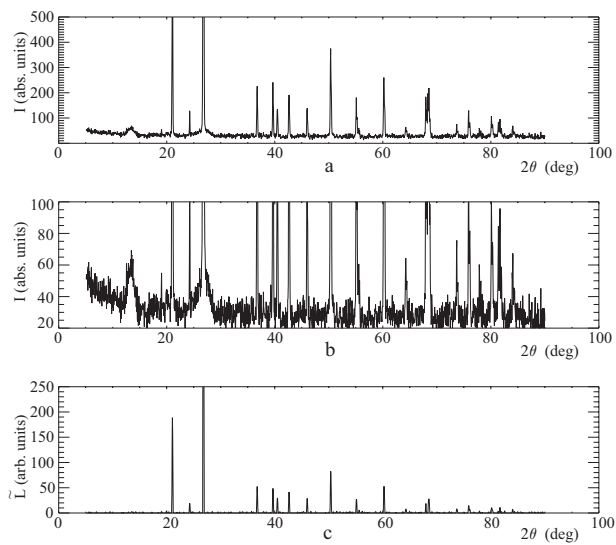}
\vskip-3mm\caption{ General appearance of the diffraction pattern
for $\alpha$-quartz ($a$); the same, but scaled up, data ($b$); and
the correlation diffraction pattern of $\alpha$-quartz
($c$)}\label{f3}
\end{figure}

The efficiency of the developed correlation analysis method was tested by
analyzing the diffraction pattern of the reference $\alpha$-$\mathrm{SiO}%
_{\mathrm{2}}$ specimen.\,\,Fi\-gure~3,~$a$ exhibits the
corresponding diffraction pattern measured with an angle increment
of 0.05$^{\circ}$.\,\,In Fig.~\ref{f3},~$b$, the same diffraction pattern
is shown on a large scale in order to demonstrate the presence of
low-intensity lines and the back\-ground.\,\,The result of a pattern
processing using the correlation method is shown in
Fig.~3,~$c$.\,\,Here, the correlation diffraction pattern was
discriminated at the level $\delta\leq4$ (see Fig.~\ref{f2}), i.e.\,\,all
points in the correlation diffraction pattern that were less than or
equal to 4 were put equal \mbox{to zero.}

\begin{table*}[!]
\vskip4mm \noindent\caption{Values of interplanar distances, angles,
and normalized intensities for $\boldsymbol\alpha$-quartz  }
\vskip3mm\tabcolsep14.0pt
\noindent{\footnotesize\begin{tabular}{|c|c|c|c|c|c|c|c|c| }
 \hline \multicolumn{3}{|c}
{\rule{0pt}{5mm}Our experiment} & \multicolumn{3}{|c}{COD
96-101-1177 \citep{COD1177}}& \multicolumn{3}{|c|}{COD
96-153-2513 \citep{COD2513}}\\[1.5mm]
\cline{1-9}
\multicolumn{1}{|c}{\rule{0pt}{5mm}$d$,~\AA}&\multicolumn{1}{|c}{$2
\theta$, deg}&\multicolumn{1}{|c}{$\widetilde{\rm
I}$}&\multicolumn{1}{|c}{$d$,~\AA}& \multicolumn{1}{|c}{$2 \theta$,
deg}&\multicolumn{1}{|c}{$\widetilde{\rm
I}$}&\multicolumn{1}{|c}{$d$,~\AA}&
\multicolumn{1}{|c}{$2 \theta$, deg}&\multicolumn{1}{|c|}{$\widetilde{\rm I}$}\\[1.5mm]%
\cline{1-9}
\multicolumn{1}{|c}{\rule{0pt}{5mm}1}&\multicolumn{1}{|c}{2}&\multicolumn{1}{|c}{3}&
\multicolumn{1}{|c}{4}&\multicolumn{1}{|c}{5}&\multicolumn{1}{|c}{6}&
\multicolumn{1}{|c}{7}&\multicolumn{1}{|c}{8}&\multicolumn{1}{|c|}{9}\\[1.5mm]
\hline%
{} \rule{0pt}{5mm}4.2205 & 21.05 & 186.79 & 4.2435 & 20.93 & 159.35 & 4.2362 & 20.97 & 202.96 \\ 
{} 3.6703 & 24.25 & 18.90 & & & & & & \\%
{} 3.3266 & 26.80 & 1000 & 3.3366 & 26.72 & 1000 & 3.3303 & 26.77 & 1000 \\%
{} 2.4456 & 36.75 & 51.92 & 2.4500 & 36.68 & 71.99 & 2.4457 & 36.75 & 75.85 \\%
{} 2.2732 & 39.65 & 48.25 & 2.2780 & 39.56 &70.28 & 2.2734 & 39.65 & 76.93 \\%
{} 2.2274 & 40.50 & 29.69 & 2.2311 & 40.43 & 22.67 & 2.2271 & 40.51 & 34.8 \\%
{} 2.1200 & 42.65 & 41.15 & 2.1218 & 42.62 & 38.49 & 2.1181 & 42.69 & 52.94 \\%
{} 1.9731 & 46.00 & 28.55 & 1.9748 & 45.96 & 24.85 & 1.9713 & 46.04 & 30.61 \\%
{} 1.8123 & 50.35 & 81.89 & 1.8144 & 50.29 & 136.96 & 1.8109 & 50.39 & 116.8 \\%
{} 1.7957 & 50.85 & \colorbox[rgb]{0.75,0.75,0.75}{2.01} & 1.8000 & 50.72 & 1.99 & 1.7962 & 50.84 & 3.26 \\[-0.5mm]%
{} 1.6668 & 55.10 & 26.74 & 1.6683 & 55.05 & 31.64 & 1.6651 & 55.16 & 40.95 \\%
{} 1.6530 & 55.60 & 5.23 & 1.6571 & 55.45 & 14.98 & 1.6537 & 55.58 & 16.66 \\%
{} 1.6028 & 57.50 & \colorbox[rgb]{0.75,0.75,0.75}{1.38} & 1.6039 & 57.46 & 3.92 & 1.6011 & 57.57 & 2.42 \\[-0.5mm]%
{} 1.5372 & 60.20 & 51.85 & 1.5375 & 60.19 & 87.69 & 1.5348 & 60.31 & 87.55 \\%
{} 1.4482 & 64.30 & 7.41 & 1.4506 & 64.21 & 15.16 & 1.4477 & 64.35 & 17.65 \\%
{} 1.4146 & 66.05 & \colorbox[rgb]{0.75,0.75,0.75}{1.78} & 1.4145 & 66.05 & 4.36 & 1.4121 & 66.18 & 3.23 \\[-0.5mm]%
{} 1.3796 & 67.95 & 18.03 & 1.3789 & 67.98 &  43.62 & 1.3764 & 68.13 & 53.31 \\%
{}   1.3716 & 68.40 &  12.55 & 1.3726 & 68.34 & 65.99 & 1.3699 & 68.49 & 61.12 \\%
{} 1.3689 & 68.55 & 27.78 & 1.3683 & 68.58 & 31.6 & 1.3659 & 68.72 & 42.63 \\%
{} 1.2855 & 73.70 & 7.42 & 1.2865 & 73.63  & 17.36  & 1.2838 & 73.81 & 22.2 \\%
{} 1.2536 & 75.90 & 14.66 & 1.253 & 75.95 & 26.75 & 1.2507 & 76.11 & 25.81 \\%
{} 1.2405 & 76.85 & 1.71 &  &  & & & & \\%
{} 1.2264 & 77.90 & \colorbox[rgb]{0.75,0.75,0.75}{3.51} & 1.2250 & 78.00 & 16.37 & 1.2229 & 78.16 & 12.3 \\[-0.5mm]%
{} 1.1975 & 80.15 & 10.46 & 1.1975 & 80.15 &25.07 & 1.1952 & 80.34 & 28.07 \\%
{} 1.1950 & 80.35 & 4.11 & 1.1946 & 80.38 & 8.07 & 1.1926 & 80.55 &  8.3 \\%
{} 1.1828 & 81.35 & \colorbox[rgb]{0.75,0.75,0.75}{3.60} & 1.1824 & 81.39 &21.89 & 1.1800 & 81.59 & 21.85 \\[-0.5mm]%
{} 1.1787 & 81.70 & 11.00 & 1.1769 & 81.85 & 24.01 & 1.1749 & 82.02 & 26.87\\%
{} 1.1757 & 81.95 & \colorbox[rgb]{0.75,0.75,0.75}{2.36} & & & & & & \\[-0.5mm]%
{} 1.1483 & 84.35 & \colorbox[rgb]{0.75,0.75,0.75}{2.62} & 1.1499 & 84.19 & 18.56 & 1.1479 & 84.38 & 14.06 \\[-0.5mm]%
{} 1.1417 & 84.95 & \colorbox[rgb]{0.75,0.75,0.75}{2.17}& 1.1390 & 85.19 & 2.69 & 1.1367 & 85.41 & 2.61 \\[-0.5mm]%
{} 1.1189 & 87.10 & \colorbox[rgb]{0.75,0.75,0.75}{1.76} & 1.1156 & 87.43 & 0.26 & 1.1135 & 87.63 & 0.22 \\[-0.5mm]%
{} 1.1143 & 87.55 & \colorbox[rgb]{0.75,0.75,0.75}{2.25} & 1.1122 & 87.76 & 2.4 & 1.1101 & 87.97 & 2.55  \\[2mm]%
\hline
\end{tabular}}
\end{table*}

The values for the angles and the relevant interplanar distances for
$\alpha $-quartz from the COD database \citep{COD}, as well as the
values obtained from our experimental data using the correlation
method, are quoted in Table.\,\,These results make it possible to
draw the following conclusions.

\raisebox{0.5mm}[0cm][0cm]{{\footnotesize$\bullet$}}\,\,The values
of angles and relative intensities for lines taken from the COD
database (see Table, columns 5--6 and 8--9) have the corresponding
values in our experimental data (Table, columns 2--3) processed with
the help of the described method.

\raisebox{0.5mm}[0cm][0cm]{{\footnotesize$\bullet$}}\,\,Close lines
at angles of 50$^{\circ}$, 55$^{\circ}$, 75.9$^{\circ}$,
81.5$^{\circ}$, 84.4$^{\circ}$, and 87.5$^{\circ}$ have been
resolved and identified, which testifies to a high resolution of the
method.

\raisebox{0.5mm}[0cm][0cm]{{\footnotesize$\bullet$}}\,\,The method
proposed in this work makes it possible to calculate the area and
intensity of lines against the \textquotedblleft
pedestal\textquotedblright\ background, as is shown in Figs.~\ref{f3},~$a$
\mbox{and~$b$.}

\section{Capabilities of the Method\\ for Determining Line Intensities.\\ A
Possibility of Using Classical Methods\\ for Evaluating the
Correlation Coefficient}

The analysis made above testifies that the developed correlation
method turned out rather sensitive, when searching for an answer to
the question: Does a selected angle interval in the diffraction
pattern contain a line?\,\,The questions of current concern are as
follows: Is it possible to draw quantitative conclusions from the
correlation diffraction pattern about the correlation between the
areas of the corresponding lines in the initial and correlation
diffraction patterns?\,\,Does the correlation diffraction pattern
only reflects the degree of correlation between the real and test
lines, or it also reflects their intensities?

\begin{figure*}
\vskip1mm
\includegraphics[width=11.7cm]{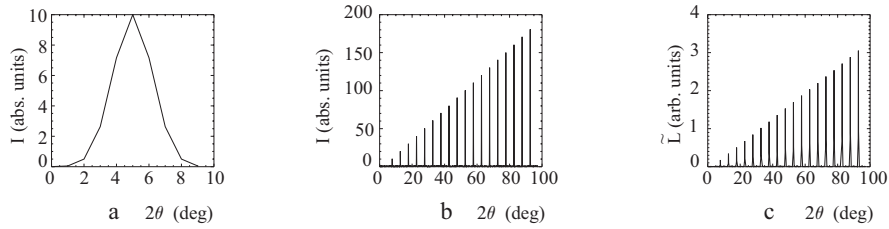}
\vskip-2mm\parbox{11.7cm}{\caption{ Model or test line ($a$), model
diffraction pattern formed from model lines with increasing
intensities ($b$), and the corresponding correlation diffraction
pattern ($c$)\label{f4}}}\vspace*{-2mm}
\end{figure*}

At first glance, the standard Pearson correlation coefficient $q$
given by the expression \citep{Pirson}
\begin{equation}
q=\frac{{\dfrac{1}{n}\sum\limits_{i=1}^{n}{(z_{i}-\bar{z})(y_{i}-\bar{y})}}%
}{{\sigma_{y}\sigma_{z}}}\label{4}%
\end{equation}
seems to be applicable in this case.\,\,When analyzing the model
samples (Fig.~\ref{f4}), this coefficient was calculated in parallel with
the coefficient $L$ at each permutation, and the value of $q$ at the
histogram maximum was taken as the correlation magnitude.\,\,In such
a manner, the most probable value of the correlation coefficient was
obtained (see, e.g., work \citep{Tanizaki}).

The both considered methods used in the calculation of the
correlation coefficient give information about the correlation
degree between two samples.\,\,However, there is a principal
difference between the Pearson correlation coefficient and the
result of the permutation test.\,\,The Pearson correlation
coefficient is introduced by expression (\ref{4}), and the value of
$q$ changes within the interval (0, 1).\,\,At the same time, the
correlation value obtained from the permutation test is proportional
to expression (\ref{3}).\,\,The\-re\-fore, if the line intensity
changes, which is equivalent to the multiplication of a sample $Y$
by a definite coefficient, expression (\ref{4}) remains unchanged,
whereas expression (\ref{3}) increases by the corresponding factor.

\begin{figure}
\vskip1mm
\includegraphics[width=5.8cm]{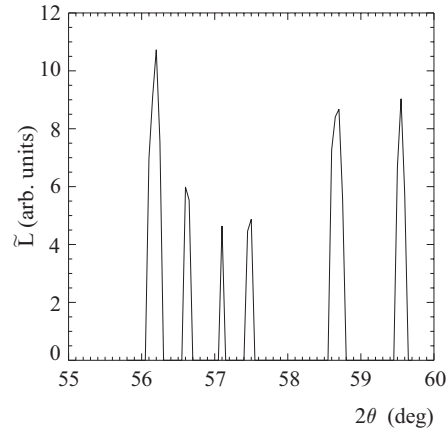}
\vskip-3mm\caption{Example of a section in the correlation
diffraction pattern, which contains additional lines, whose
intensities are higher than the intensity of the tabular line at{
$2\theta=57.46^{\circ}$$\div$$57.57^{\circ}$} }\label{f5}
\end{figure}
\begin{figure}
\vskip1mm
\includegraphics[width=6.5cm]{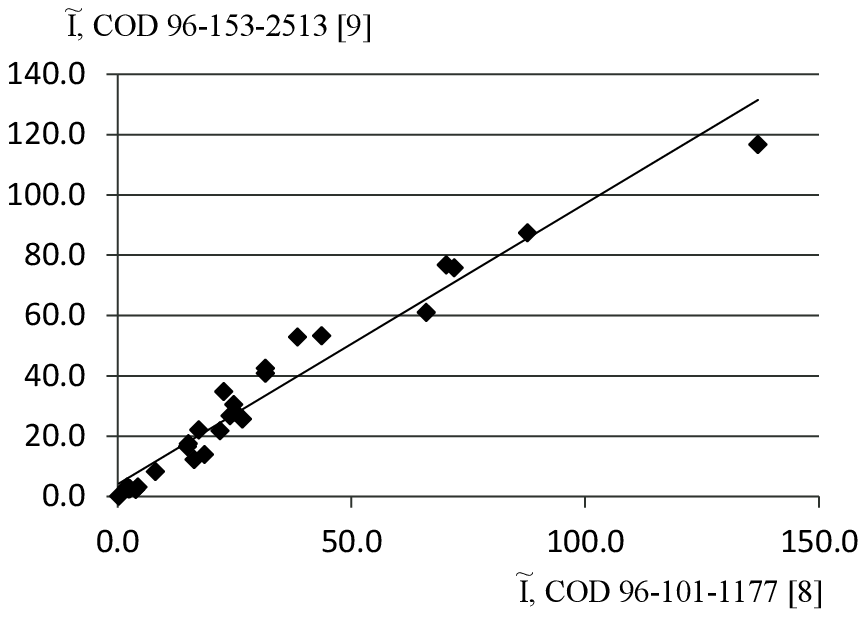}
\vskip-3mm\caption{Illustration of deviations from the linear
relation between the line intensities for $\alpha$-quartz taken from
the COD database {\citep{COD}}}\label{f6}
\end{figure}

\begin{figure*}
\vskip1mm
\includegraphics[width=11.4cm]{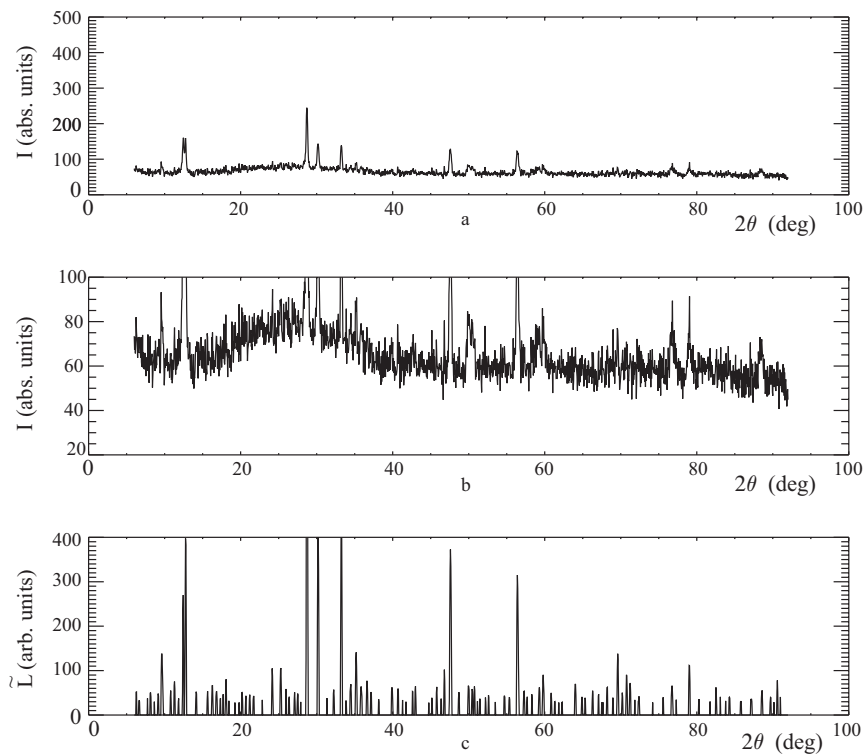}
\vskip-2mm\parbox{11.4cm}{\caption{ General appearance of a
brown-ceramic diffraction pattern ($a$); the same, but scaled up,
data ($b$); and the correlation diffraction pattern of brown
ceramics ($c$)\label{f7}}}\vspace*{-1mm}
\end{figure*}

This statement was verified by means of a direct simulation.\,\,A
model diffraction pattern was generated (Fig.~\ref{f4},~$a$), which
included a noise background and a series of Gaussian lines with
gradually increasing amplitudes (Fig.~\ref{f4},~$b$).\,\,The result of the
correlation processing of this signal is shown in
Fig.~\ref{f4},~$c$.\,\,One can see that the intensity of correlation lines
also increases linearly, which means that there is a linear
relationship between the intensities of lines in the original
diffraction pattern and the intensities of the corresponding
correlation lines.\,\,From whence the following conclusion can be
drawn: the amplitude of lines in the correlation diffraction pattern
reflects the amplitude of lines in the initial diffraction pattern,
but it is obtained in terms of some relative units, because it
depends, e.g., on the amplitude of a test line.\,\,The\-re\-fore,
for the comparison with other data to be possible, the procedure of
obtaining the correlation diffraction pattern has to be
calibrated.\,\,The required normalizing factor can be obtained,
e.g., by analyzing the diffraction pattern for $\alpha$-quartz and
equating the amplitude of the maximum correlation diffraction line
to 1000, as it is customary to do in databases (see Table).

Figure~\ref{f4} testifies that there is a linear relationship between the
intensities of the experimental and correlation diffraction
lines.\,\,The\-re\-fore, all lines in the correlation diffraction
pattern (after the discrimination of low-intensity lines) are real
by definition, including those detected in noise.\,\,The\-re\-fore,
after the normalization of the correlation diffraction pattern, it
can be argued that the relative intensity of experimental
diffraction lines (it is determined with a certain error according
to the noise level) will be more correct in the correlation
diffraction pattern.

This issue can be discussed by analyzing Table, which contains data
of various experiments.\,\,One can see that there are some
differences between the tabulated data themselves and the data
obtained in this work.\,\,The intensities of some lines are very low
(the corresponding cells in Table are colored).\,\,At the same time,
proceeding from our data, we obtain a paradox.\,\,For
instance, the line indicated in the database at $2\theta=57.46^{\circ}%
$$\div$$57.57^{\circ}$ is also visible in our correlation diffraction pattern
(Fig.~\ref{f5}).\,\,Ho\-we\-ver, the same angle interval includes some more
low-intensity lines that are not contained in the tabular data,
although their intensities
are higher than that of the tabular line at $2\theta=57.46^{\circ}%
$$\div$$57.57^{\circ}$.\,\,This fact means that either several additional
lines belonging to $\alpha$-quartz revealed themselves in the
correlation spectrum in Fig.~\ref{f5} or all lines shown in Fig.~\ref{f5},
including the tabular line at
$2\theta=57.46^{\circ}$$\div$$57.57^{\circ}$, are \textquotedblleft
noise\textquotedblright.\,\,The latter may arise as a result of the
imperfect character of the reference specimens (both ours and those,
the results from which were included into the indicated
database).\,\,From whence, it follows that the lines, whose
intensities are at the unity level may be real, but not associated
with the perfect structure of the crystal lattice in the examined
homogeneous material.\,\,A decision of whether they should be taken
into account, when identifying the phases by comparing with the
tabulated angle values, should have a special substantiation.

On the basis of the data quoted in Table, a plot illustrating the
relationship between the tabular values for line intensities taken
from the COD database \citep{COD} can be drawn (see
Fig.~\ref{f6}).\,\,Ex\-pec\-ted\-ly, there is a cretain discrepancy between
those values, which is a result of the available experimental
accuracy of diffraction patterns.\,\,From the results of model
calculations (Fig.~\ref{f4}), it follows that there is a linear
relationship between the correlation and model intensities (a plot
analogous to that in Fig.~\ref{f6} cannot be drawn for experimental data,
because there is no way to independently evaluate the intensities of
experimental lines, which are determined exactly in model
calculations).\,\,The\-re\-fore, it is reasonable to assume that the
correlation intensities are more accurate, so that it is better to
use the correlation intensities given in column 3 of Table, when
estimating the relationships between the $\alpha$-quartz lines.

On the basis of our researches, we may conclude that the method
developed for the correlation analysis of diffraction patterns gives
reliable results.\,\,This conclusion allowed this method to be
applied to the analysis of LLFCMs that look like (and
correspondingly dubbed) ''brown ceramics''.

In Fig.~\ref{f7},~$a$, the diffraction pattern for a specimen of LLFCM
brown ceramics is shown.\,\,The diffraction pattern of this material
looks like a noise track with six intense lines and a few broadened
ones, e.g., at $\theta\approx 50^{\circ}$ and 60$^{\circ}$ (see
Fig.~\ref{f7},~$b$).\,\,The available data make it possible to reliably
identify (on the basis of six lines) and to determine the content of
only one phase: uranium oxide UO$_{2.234}$ (4.5--5.5 wt.\%).\,\,The
presence of the following phases are supposed: cubic zirconium oxide
$\mathrm{ZrO}_{2}$ (five lines), orthorhombic SiO$_{2}$ (one line),
and uranium silicate USiO$_{7}$ (one line).\,\,For all indicated
crystalline phases, their structural (non-chemical) formulas in
accordance with the COD database were used.\,\,At first glance, the
processing of diffraction data in the framework of our correlation
analysis method seems to give a similar picture (Fig.~\ref{f7},~$b$): six
intensive lines and a noise track in the form of low-intensity
lines.

However, one should bear in mind that \textquotedblleft
noises\textquotedblright\ have already been removed from the
correlation diffraction pattern.\,\,The\-re\-fore, according to the
discussion above, all lines in the correlation diffraction pattern
should be considered as real.\,\,A comparison of those lines with
the X-ray diffraction database \citep{COD} brought about an
unexpected result.\,\,By analyzing this diffraction pattern, we can
not only confirm the identification of uranium oxide UO$_{2{,}234}$,
but also reliably identify (on the basis of more than six lines) and
determine the content of cubic zirconium oxide ZrO$_{2}$
(2--3~wt.\%), orthorhombic silicon oxide SiO$_{2}$ (3--5~wt.\%), and
uranium silicate USiO$_{7}$ (3--4~wt.\%).\,\,The correlation
diffraction pattern allowed the identification of phases with the
content not only at a level of 1~wt.\% ($\mathrm{Al}_{0.32}\mathrm{Si}%
_{0.68}\mathrm{O}_{2}$), but also lower than 1~wt.\%, namely,
triclinic silicon oxide (0.4--0.5$\mathrm{~wt.\%}$), cubic silicon
oxide SiO$_{2}$ (0.2--0.4$\mathrm{~wt.\%}$), calcium silicate
CaSiO$_{3}$ (0.15--0.3~wt.\%), and calcium silicate
$\mathrm{Ca}_{3}\mathrm{Si}_{2}\mathrm{O}_{6}$
(0.1--0.2$\mathrm{~wt.\%}$).

We note that the high sensitivity and accuracy of the method poses a
new problem. The existence of a great number of low-intense lines at
the analysis ``by hands'' increases the error of determination of
their relative contribution (errors can increase by several times).
The question about the development of a method of reliable
evaluation of the content of phases in such situation will be
considered in the subsequent works.

From the viewpoint of the X-ray diffractometry practice, this result
of the crystalline phase identification by processing the
diffraction patterns can be regarded as a very
good~one.\vspace*{-2mm}

\section{Conclusions}

A method to detect low-intensity X-ray lines has been developed for
the problems of X-ray phase analysis of materials that contain many
(more than eight) low-content (from several tenths of wt.\% to a few
wt.\%) crystalline phases.\,\,The method is based on the calculation
of correlations with the use of the computer statistics approaches.

The application of this method to reference specimens showed not
only its efficiency, but also its advantages, when determining the
line intensities.\,\,In particular, in many cases, when the database
contains only a qualitative indication that the intensity of
corresponding lines is actually lower than the identification level,
the developed method allowed the corresponding intensity value to be
assigned to each detected line (see Table).

The application of the developed correlation method to the analysis
of the diffraction patterns of specimens containing phases with a
content lower than 0.1~wt.\% showed that such phases can be reliably
identified.\,\,The determination of the lower sensitivity limit of
the method is a task for the further research.

\vskip2mm {\it The work was sponsored in the framework of the budget
theme of the National Academy of Sciences of Ukraine
(No.~0117U002636)}.

\rezume{%
А.Д.\,Скорбун, С.В.\,Габєлков, І.В.\,Жиганюк,\\ В.Г.\,Кудлай,
П.Є.\,Пархомчук, С.О.\,Чиколовець}{МЕТОД ОБРОБКИ ДАНИХ
РЕНТГЕНІВСЬКОЇ\\ ДИФРАКЦІЇ ДЛЯ БАГАТОФАЗНИХ МАТЕРІАЛІВ\\ З НИЗЬКИМ
ВМІСТОМ ФАЗ} {Під час дослідження методом рентгенівського фазового
аналізу аморфних, скляних та склокерамічних матеріалів, які
практично завжди мають у своєму складі  значну кількість
кристалічних фаз (більше восьми) з їх вмістом від декількох \% мас.
до декількох десятих~-- сотих \% мас., виникають труднощі з
визначенням фазового складу. Розроблено метод аналізу дифрактограм
таких матеріалів, коли на дифрактограмах спостерігаються
рентгенівські лінії, інтенсивність яких знаходиться на рівні шумів.
Виявлення ліній базується на використанні пошуку кореляцій між
експериментальною і тестовою лініями і перевірці збігів за допомогою
статистичних методів аналізу (комп'ютерна статистика). Метод
перевірено на зразках $\alpha$-кварцу, який часто використовується
як еталон, та застосовано до аналізу лавоподібних паливовмісних
матеріалів Чорнобильської АЕС. Показано, що розроблена методика
дозволяє ідентифікувати рентгенівські лінії при вмісті окремих фаз
до $0{,}1$~\% мас.,  а також істотно підвищує можливість кількісно
визначати вміст фази за лініями з слабкою інтенсивністю.}

\end{document}